\documentclass[nofootinbib,article,
 reprint,
 amsmath,amssymb,
 aps,
]{revtex4-1}

\usepackage{filecontents}
\usepackage{graphicx}
\usepackage{dcolumn}
\usepackage{bm}
\usepackage{subfigure}
\usepackage{graphicx}

\begin{document}

\preprint{APS/123-QED}

\title{Finite Temperature Phase Diagrams of a Two-band Model of Superconductivity}


\author{Heron Caldas}
\author{A. Celes}
\affiliation{Departamento de Ci\^{e}ncias Naturais, Universidade Federal de S\~{a}o Jo\~{a}o Del Rei, Pra\c{c}a Dom Helv\'{e}cio 74, 36301-160, S\~{a}o Jo\~{a}o Del Rei, MG, Brazil\\}

\author{David Nozadze}
\affiliation{Cisco Systems, Inc., San Jose, CA, 95134, USA}

\date{\today}

\begin{abstract}
We explore the temperature effects in the superconducting phases of a hybridized two-band system. We show that for zero hybridization between the bands, there are two different critical temperatures. However, for any finite hybridization there is only one critical temperature at which the two gaps vanish simultaneously. We construct the phase diagrams of the critical temperature versus hybridization parameter $\alpha$ and critical temperature versus critical chemical potential asymmetry $\delta \mu$ between the bands, identifying the superconductor and normal phases in the system. We find an interesting reentrant behavior in the superconducting phase as the parameters $\alpha$ or $\delta \mu$, which drive the phase transitions, increase. We also find that for optimal values of both $\alpha$ and $\delta \mu$ there is a significant enhancement of the critical temperature of the model.

\end{abstract}

\pacs{Valid PACS appear here}
\maketitle


\section{\label{intro}Introduction}

Magnesium diboride ($\rm{MgB_{2 }}$) is a simple and, at the same time, unusual superconductor. Experimental measurements have indicated that $\rm{MgB_{2 }}$ has two distinct superconducting gaps~\cite{exp1,exp2,exp3,exp4,exp5,exp6,exp7,exp8}, but only one critical temperature ($T_c$). With a $T_c \sim 40 ~ \rm{K}$~\cite{exp1} this metallic compound has the highest known critical temperature at {\it ambient} pressure amongst conventional superconductors.

Hybridization i.e., the mixing of atomic orbitals, plays an important role in the physics of multiband superconductors (see, for instance~\cite{hybri1,hybri2,hybri3,hybri4,hybri5,hybri6,hybri7,hybri8}).  This seems to be also the case for $\rm{MgB_{2 }}$. Indeed, as shown in Ref.~\cite{Choi}, the $\rm{MgB_{2 }}$ Fermi surface (FS) is determined by three orbitals, but only two different energy gaps are experimentally detected. This happens because two of the three orbitals hybridize among themselves and determine one single band, responsible for a large superconducting gap on the $\sigma$ FS, while the non-hybridized orbital determines a smaller superconducting gap at the FS of the $\pi$ band.

As pointed out in Ref.~\cite{Komendova}, besides $\rm{MgB_{2 }}$ the importance of multiband superconductivity has also been suggested to other materials as, for instance, simple metals~\cite{Shen,Boaknin}, and heavy fermion compounds~\cite{Stewart,Jourdan,Rouke,Seyfarth,Hill}.

The hybridization among orbitals can be symmetric or antisymmetric under inversion symmetry. It has been shown that symmetric ($k$-independent) hybridization acts in detriment of intra-band superconductivity~\cite{sym1,sym2}. On the other hand, antisymmetric ($k$-dependent) hybridization enhances superconductivity~\cite{anti1}. It has been considered recently the cases at which two bands are formed by electronic orbitals with angular momentum, such that, the k-dependent hybridization $V(k)$ between them can be symmetric or antisymmetric~\cite{Twoband}. Only intra-band attractive interactions have been taken into account in these two bands and the appearance of induced inter-band pairing gaps were investigated. It was shown that these inter-band superconducting orderings are induced even in the total absence of attractive interaction between the two bands, which turns out to be completely dependent on the hybridization between them. For the case of antisymmetric hybridization, which causes an odd-parity mixing between the $a$ and $b$ bands, the induced inter-band pairing gap that emerges in the hybridized bands has $p$-wave symmetry~\cite{Twoband}.

In this work we study the temperature effects on intra-band pairing gaps under the influence of the hybridization of two single bands, say $a$ and $b$. We consider superconducting interactions only inside each band, resulting in intra-band pairing gaps $\Delta_a$ and $\Delta_b$, respectively, in these bands. We take into account symmetric and antisymmetric $V(k)$. We find how the critical temperature of the hybridized system depend on the strength of the hybridization and on the particles chemical potential asymmetry. Based on these results, we construct the respective phase diagrams of the model and find an intriguing reentrant behavior. For optimal values of both $\alpha$ and $\delta \mu$ we find a significant enhancement of the critical temperature of the model.

The paper is organized as follows: In Sec.~\ref{model} we introduce the generic model Hamiltonian describing the two-band system. In Sec.~\ref{gap1} we obtain the diagonalized Hamiltonian and the grand thermodynamic potential of the model from which the gap equations for symmetric hybridization are derived. From the self-consistent solutions of the gap equations the critical temperatures and chemical potential asymmetries are obtained and from these results the corresponding phase diagrams are constructed. Sec.~\ref{gap2} contains the same investigations of the previous section, but for antisymmetric hybridization. We conclude in Sec.~\ref{Co}.

\section{\label{model} Model Hamiltonian}

The Hamiltonian describing a 3D effective generic two-band supercondutor model in second quantization is given by~\cite{Twoband}

\begin{align}
\label{eq:Hab}
H_{ab}  & =  \sum_{{\bf k}\sigma} \varepsilon_a ({\bf k}) a^\dagger_{{\bf k}\sigma} a_{{\bf k}\sigma} + \sum_{\bf k} \Delta_a({\bf k}) a^\dagger_{{\bf k}\uparrow} a^\dagger_{-{\bf k}\downarrow} +{\rm h.c.} \nonumber  \\
& +\sum_{{\bf k}\sigma} \varepsilon_b ({\bf k}) b^\dagger_{{\bf k}\sigma} b_{{\bf k}\sigma} +  \sum_{\bf k} \Delta_b({\bf k}) b^\dagger_{{\bf k}\uparrow} b^\dagger_{-{\bf k}\downarrow} +{\rm h.c.}  \nonumber \\ 
& +  \sum_{{\bf k}\sigma} V({\bf k}) a^\dagger_{{\bf k}\sigma} b_{{\bf k}\sigma} + {\rm h.c.} - \frac{ |\Delta_a|^{2}}{g_a} - \frac{ |\Delta_b|^{2}}{g_b},
\end{align}
where the operator $a^\dagger_{{\bf k}\sigma}$ creates an electron in band $a$ with momentum ${\bf k}$ and spin $\sigma$, and similarly for band $b$. The kinetic energy is given by the band dispersions $\varepsilon_{\eta}= \frac{k^2}{2m_\eta}-\mu_\eta$, where $\eta=a,b$, with $m_\eta$ denoting the electron mass. $g_a$ and $g_b$ are the coupling constants of the electrons in the respective bands. 

In order to guarantee homogeneous equilibrium, we have set the same chemical potential $\mu$ for all electrons occupying different bands. Thus, we can define an efective chemical potential of the electrons in bands $a$ and $b$ as $\mu_\eta=\mu+E_\eta$. The constants $E_\eta$ are the bottom of the specific band $\eta$.

Although the hybridization $V({\bf k}) \equiv V_k$, or single-particle band scattering, has been considered several times in the literature as, for example, in Refs.~\cite{Japiassu,Moreo,essential,Annals1}, it deserves some comments. As discussed in Ref.~\cite{Annals1} (and references therein), the parity of the hybridization depends on the difference of the angular momenta $\Delta l = l' - l$ of the mixed electronic orbitals. That is, $V_k$ has even parity if $\Delta l$ is an even number, and odd parity if $\Delta l$ is an odd number. As a result, if the neighbor mixed orbitals have angular momentum $l$ and $l'=l+1$ the odd-parity has necessarily to be taken into account.

The original (i.e., before $V_k$ be ``turned on'') superconducting intra-band {\it mean-field} order parameters are $\Delta_{a,b}$. We will assume here, as often, spin-singlet, regular $s$-wave superconducting states. However, the methods performed here could be generalized to other kinds of intra-band pairings.

\section{\label{gap1} Obtention of the Gap Equations for Symmetric Hybridization $V(k)$}

The Hamiltonian in Eq.~(\ref{eq:Hab}) can be rewritten in the basis~$\Psi_{{\bf k}} = (a_{\textbf{k},\uparrow}, b_{\textbf{k}, \uparrow},
a_{-\textbf{k},\downarrow}^{\dag}, b_{-\textbf{k},\downarrow}^{\dag})^{T}$ as:

\begin{eqnarray}
H=\frac{1}{2} \sum_{{\bf k}}\Psi_{{\bf k}}^{\dag}\mathcal{H}({\bf k})\Psi_{{\bf k}}
 + 2 \sum_{\textbf{k}} \varepsilon_{+}({\bf k}) - \frac{ |\Delta_a|^{2}}{g_a} - \frac{ |\Delta_b|^{2}}{g_b}, 
 \label{6}
\end{eqnarray}
with

\begin{eqnarray}
\mathcal{H}({\bf k})=\left(
  \begin{array}{cccc}
    \varepsilon_a ({\bf k}) & V_{{\bf{k}}}^{*}  & -\Delta_{a}^{*} &  0  \\
 V_{{\bf{k}}} & \varepsilon_b ({\bf k}) & 0 & -\Delta_{b}^{*} \\
   - \Delta_a & 0 & -\varepsilon_a ({\bf k}) &  -V_{{\bf{k}}} \\
    0 & -\Delta_b & -V_{{\bf{k}}}^{*} &  -\varepsilon_b ({\bf k})  \\
  \end{array}
\right). 
\label{7}
\end{eqnarray}

This Hamiltonian can be diagonalized as

\begin{eqnarray}
H&=&\sum_{\textbf{k},s=1,2} E_{\textbf{k},s}\alpha_{\textbf{k},s}^{\dag}\alpha_{\textbf{k},s} \\
\nonumber
&+& \sum_{\textbf{k},s=1,2} (2\varepsilon_{+}({\bf k}) -E_{\textbf{k},s}) - \frac{|\Delta_a|^{2}}{g_a} - \frac{|\Delta_b|^{2}}{g_b}, 
\label{8}
\end{eqnarray}
where $\varepsilon_{+}({\bf k}) \equiv \frac{\varepsilon_a ({\bf k})+\varepsilon_b ({\bf k})}{2}$, and $\alpha_{\textbf{k}, 1,2}^{\dag}(\alpha_{\textbf{k}, 1,2})$ is the creation(annihilation) operator for the quasiparticles with excitation spectra

\begin{widetext}
\begin{eqnarray}
E_{{\bf{k}}, 1,2}\!=\!\frac{1}{2} \sqrt{ 2  E_{{\bf{k}}}^{2} \!+\! 4 |V_{{\bf{k}}}|^2 \!\pm \!2 \sqrt{\! \big( |\Delta_a|^{2}\!-\!|\Delta_b|^{2} \!+\! \varepsilon_a ({\bf k})^2 \!-\! \varepsilon_b ({\bf k})^2 \big)^2\!+\! 4|V_{{\bf{k}}}|^2\Big[\big(\varepsilon_a ({\bf k}) \!+\! \varepsilon_b ({\bf k}) \big)^2 \!+\! |\Delta_{a}|^{2} \!+\! |\Delta_{b}|^{2}\Big] \!-\! 8 Re[\Delta_{a}\Delta_{b}^{*} V_{{\bf{k}}}^{2}]} }, \nonumber\\
\end{eqnarray}
where we have defined $E_{{\bf{k}}}^{2} \equiv  |\Delta_a|^{2}+|\Delta_b|^{2} + \varepsilon_a ({\bf k})^2 + \varepsilon_b ({\bf k})^2 $. The other two quasiparticles energies are $E_{{\bf{k}}, 3}=-E_{{\bf{k}}, 1}$ and $E_{{\bf{k}}, 4}=-E_{{\bf{k}}, 2}$. Up to here the symmetric $V_{{\bf{k}}}$ can be real or complex. If we take the order parameters and the hybridization as real terms, the previous equation can be simplified as

\begin{eqnarray}
E_{{\bf{k}}, 1,2}=\frac{1}{2} \sqrt{ 2  E_{{\bf{k}}}^{2} + 4 V_{{\bf{k}}}^2 \pm 2 \sqrt{ \big( \Delta_a^{2}-\Delta_b^{2} + \varepsilon_a ({\bf k})^2 - \varepsilon_b ({\bf k})^2 \big)^2+ 4V_{{\bf{k}}}^2\Big[\big(\varepsilon_a ({\bf k}) + \varepsilon_b ({\bf k}) \big)^2 + \big(\Delta_{a} - \Delta_{b}\big)^{2} \Big] } }. \nonumber\\
\label{Energias}
\end{eqnarray}
\end{widetext}
 
 \subsection{\label{TP} The Grand Thermodynamic Potential for Symmetric Hybridization $V(k)$}

It is straightforward to write down the grand thermodynamic potential $\Omega = -  \text{Tr}\ln[e^{-\beta H}]$, where $\beta=1/(k_{B} T)$, at finite temperature,

\begin{eqnarray}
\label{poteff1}
\Omega &=&
\frac{1}{2}\sum_{\textbf{k},s=1,2} \left[\varepsilon_{+}({\bf k})-E_{\textbf{k},s} -\frac{2}{\beta} \ln(1+e^{-\beta E_{\textbf{k},s}}) \right] \nonumber\\
&-& \frac{\Delta_a^{2}}{g_a} - \frac{\Delta_b^{2}}{g_b},
\end{eqnarray}
from which all quantities of interest can be obtained. The quasiparticles excitation energies $E_{\textbf{k},s}$ above are given by Eq.~(\ref{Energias}).

Minimization of $\Omega$ in Eq.~(\ref{poteff1}) with respect to the gaps $\Delta_a$ and $\Delta_b$ respectively, gives

\begin{eqnarray}
\label{gapequaG1}
&&\frac{4\Delta_a}{g_a} = \sum_{\textbf{k}} \! \left[\tanh \left(\frac{E_{\textbf{k},1}}{2T} \right) \frac{\partial E_{\textbf{k},1}}{\partial \Delta_a} \!+\! \tanh\left(\frac{E_{\textbf{k},2}}{2T}\right) \frac{\partial E_{\textbf{k},2}}{\partial \Delta_a}  \right], \nonumber\\
\end{eqnarray}
and
\begin{eqnarray}
\label{gapequaG2}
&&\frac{4\Delta_b}{g_b} \!=\! \sum_{\textbf{k}} \! \left[\tanh \left(\frac{E_{\textbf{k},1}}{2T} \right) \frac{\partial E_{\textbf{k},1}}{\partial \Delta_b} \!+ \! \tanh \left( \frac{E_{\textbf{k},2}}{2T} \right) \frac{\partial E_{\textbf{k},2}}{\partial \Delta_b}  \right]. \nonumber\\
\end{eqnarray}

The equations above are developed in Appendix~\ref{GE1}, where we defined $\frac{1}{\lambda_a} = {\cal{F}}$, $\frac{1}{\lambda_b} = {\cal{G}}$, with $\lambda_a \equiv g_a \rho(0)$, $\lambda_b \equiv g_b \rho(0)$. Here $\rho(0) = \frac{m}{2 \pi^2} k_F$ is the density of states at Fermi level, and $k_F=\sqrt{2m \bar \mu}$ is the Fermi momentum. We have also defined $\bar \mu = (\mu_b + \mu_a)/2$, $\delta \mu = (\mu_b - \mu_a)/2$, and $\xi = \frac{k^2}{2m} - \bar \mu $.

We consider in this section a symmetric hybridization $V_k = \gamma k^2$ that with the above change of variables $\to$ $V_{\xi} = 2m \gamma (\bar \mu + \xi)$. In order to make the subsequent numerical analysis, we define now the non-dimensional variables $x=\xi/E_F$, $\tilde T = T/E_F$, $\tilde \Delta_{a,b} = \Delta_{a,b}/E_F$, $\tilde \delta \mu = \delta \mu/E_F$, $\tilde \mu = \bar \mu/E_F$, and $\tilde V_x = V_\xi/E_F = 2m \gamma (\tilde \mu +x)$, where $E_F = k_F^2/2m$ is the Fermi energy. Defining also a non-dimensional (for the case of the symmetric hybridization we are considering) hybridization parameter $\alpha \equiv 2 m \gamma$, we have $\tilde V_x^2 = \alpha^2 (\tilde \mu +x)^2$. Thus we can write

\begin{eqnarray}
\label{PreGL3x}
{\cal{F}} &=& \frac{1}{4} \int_0^{\bar \omega} d x \bigg\{ \tanh \left(\frac{E_{x,1}}{2\tilde T} \right) \frac{1}{E_{x,1}} \Bigg[ 1   \\
\nonumber
&+& \left. \frac{1}{E(x)} \Big[ \tilde \Delta_a^2 - \tilde \Delta_b^2 +4 \tilde \delta \mu x + 2 V_{{x}}^2 \frac{\left(\tilde \Delta_a -\tilde \Delta_b \right)}{\tilde \Delta_a} \Big] \right]\bigg\} \\
\nonumber
&+& \frac{1}{4} \int_0^{\bar \omega} d x \bigg\{  \tanh \left(\frac{E_{x,2}}{2\tilde T} \right) \frac{1}{E_{x,2}} \Bigg[ 1   \\
\nonumber
&-& \left. \frac{1}{E(x)} \Big[ \tilde \Delta_a^2 - \tilde \Delta_b^2 + 4 \tilde \delta \mu x  + 2 V_{{x}}^2 \frac{\left(\tilde \Delta_a - \tilde \Delta_b \right)}{\tilde \Delta_a} \Big] \right] \bigg\},
\end{eqnarray}
and

\begin{eqnarray}
\label{PreGL4x}
{\cal{G}}& =& \frac{1}{4} \int_0^{\bar \omega} d x \bigg\{ \tanh \left(\frac{E_{x,1}}{2 \tilde T} \right) \frac{1}{E_{x,1}} \Bigg[ 1  \\
\nonumber
&-& \left. \frac{1}{E(x)} \Big[  \tilde \Delta_a^2 - \tilde \Delta_b^2 + 4 \delta \mu x + 2 V_{{x}}^2 \frac{\left(\tilde \Delta_a - \tilde \Delta_b \right)}{\tilde \Delta_b} \Big] \right]\bigg\} \\
\nonumber
&+& \frac{1}{4} \int_0^{\bar \omega} d x \bigg\{ \tanh \left(\frac{E_{x,2}}{2 \tilde T} \right) \frac{1}{E_{x,2}} \Bigg[ 1   \\
\nonumber
&+& \left. \frac{1}{E(x)} \Big[  \tilde \Delta_a^2 - \tilde \Delta_b^2 + 4 \delta \mu x + 2 V_{{x}}^2 \frac{\left( \tilde \Delta_a - \tilde \Delta_b \right)}{\tilde \Delta_b} \Big] \right] \bigg\},
\end{eqnarray}
where $\bar \omega=\omega/E_F$, $E_{{x}, 1,2}=\frac{1}{2} \sqrt{ 2  E_{x}^{2} + 4 \tilde V_{x}^2 \pm 2  E(x)}$, $E_{x}^{2} =  \tilde \Delta_a^{2}+\tilde \Delta_b^{2} + 2 ({x}^2+{\tilde \delta \mu}^2)$ and $E(x)=\sqrt{ \big(\tilde \Delta_a^{2}- \tilde \Delta_b^{2} + 4\tilde \delta \mu x \big)^2+ 4 \tilde V_{x}^2\Big[( 2 x )^2 + (\tilde \Delta_{a} - \tilde \Delta_{b})^{2} \Big] }$. 

\subsection{Determination of the Critical Temperatures $T_{c,a}$ and $T_{c,b}$ for Symmetric $V(k)$}

\begin{figure}
\centering
 \includegraphics[width=8cm]{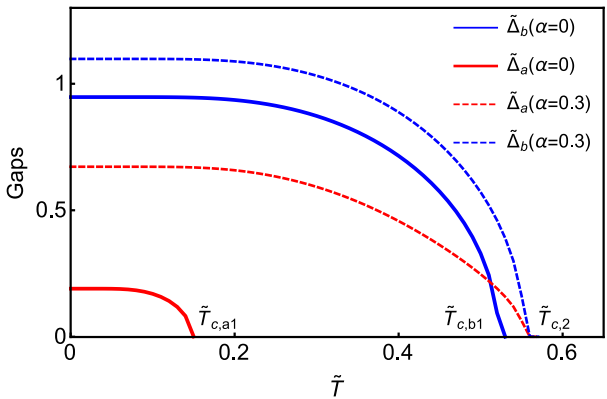}
\caption{(Color online) Gap parameters with symmetric hybridization as a function of the temperature $\tilde T$ for $\alpha=2m\gamma=0$ and $0.3$. The parameters used are $\bar \omega=10$, $\lambda_a = 0.58$, $\lambda_b = 0.6$, $\tilde \mu_a = 1.2$, and $\tilde \mu_b = 1.6.$}
\label{fig1}
\end{figure}

Given $g_a$ and $g_b$, Eqs.~(\ref{PreGL3x}) and (\ref{PreGL4x}) have to be solved self-consistently to find the gaps $\tilde \Delta_a$ and $\tilde \Delta_b$. In Fig.~(\ref{fig1}) we show the (non-dimensional) gaps $\tilde \Delta_a$ and $\tilde \Delta_b$ as a function of the normalized temperature $\tilde T$ for two different values of the hybridization strength $\alpha$. The curves are the self-consistent solutions of Eqs.~(\ref{PreGL3x}) and (\ref{PreGL4x}). At zero hybridization $V_k$ between the bands, there are two critical temperatures, $\tilde T_{c,a1}$ and $\tilde T_{c,b1}$. At finite (and symmetric) $\tilde V_k$, the smaller gap increases and acquires a new and bigger critical temperature, $\tilde T_{c,a2} =\tilde T_{c,b2} \equiv \tilde T_{c,2} $. Notice that the system still has two gaps, but there is only one critical temperature now. This is the same qualitative behavior found experimentally in $\rm{MgB_2}$, where the $\sigma$ and $\pi$ superconducting gaps vanish at the same transition temperature (see Fig. (3) of~\cite{Nagamatsu}). It worth to point out that there are some similarities with the seminal paper by Suhl, Matthias, and Walker~\cite{Suhl}. The main difference is that in~\cite{Suhl} a superconducting interband interaction is taken into account while here we consider the hybridization between the two (superconducting) single bands.

\begin{figure}
\centering
 \includegraphics[width=8cm]{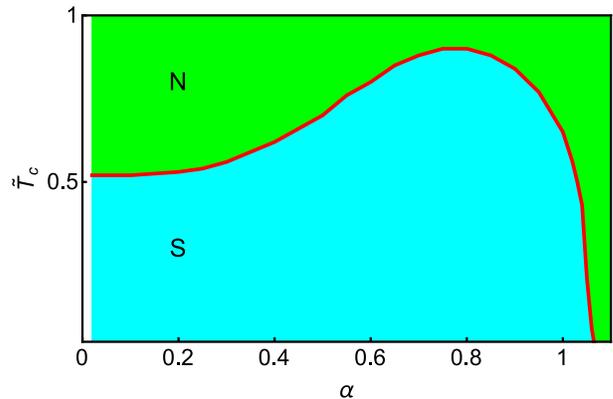}
\caption{(Color online) Phase diagram of the two-band model showing the critical temperature $\tilde T_c$ versus $\alpha$ for symmetric hybridization. Superconductivity (S) develops in the blue region and the normal (N) phase is displayed in the green region. The parameters used are $\bar \omega=10$, $\lambda_a = 0.58$, $\lambda_b = 0.6$, $\tilde \mu_a = 1.2$, and $\tilde \mu_b = 1.6.$}
\label{fig2}
\end{figure}

As we have seen in Fig.~(\ref{fig1}), for any $\alpha>0$ there is only one critical temperature for both gaps. This allows us to construct a finite temperature ``phase diagram''~\cite{Comment} of this two-band model with symmetric hybridization, which is depicted in Fig.~(\ref{fig2}). Notice that this phase diagram evidences a behavior that we could name as ``inverse reentrant behavior''\footnote{We named the behavior observed here as {\it inverse reentrant behavior} in order do not confuse with the conventional reentrant superconductivity phenomena, which happens when a continuous parameter is changed, then superconductivity is first observed, after that is destroyed by the ferromagnetic order, and later reappears.}. Beginning in any point in the normal region $N$ with $\tilde T > \tilde T_c \sim 0.51$, and going horizontally to the right as the hybridization parameter $\alpha$ increases, the system enters in the superconducting region and then enters back in the normal phase.

It is worth to point out the underlying physics for the enhancement of the critical temperatures (and consequently the pairing gaps) $\tilde T_{c,a2} =\tilde T_{c,b2} = \tilde T_{c}$ in Fig.~(\ref{fig2}), of the $a$ and $b$ bands with the increasing of the hybridization strength $\alpha$. To that aim, we can make an analogy to a two-component Fermi gas with spin-orbit coupling (SOC), although hybridization does not mix spins as SOC does. As happens in the case of SOC before this coupling to get started ~\cite{Chen}, the pairing wave function without hybridization is simply singlet. However, in the presence of SOC each energy band contains both spin-up and spin-down components~\cite{Twoband}. As a result, the pairing wave function has a more complicated structure with both singlet and triplet components~\cite{Rashba,Alicea}. Notice that in the language of the second quantization of quantum mechanics, the hybridization term in the Hamiltonian of Eq.~(\ref{eq:Hab}) is destroying an electron in one band and creating it in the other. After a critical value of $\alpha$ this process is no more energetically favorable.

The numerical calculations in Fig.~(\ref{fig2}) begin at $\alpha=0.01$ since at strict $\alpha=0$ there is no only one $\tilde T_c$, but $\tilde T_{c,a1}$ and $\tilde T_{c,b1}$ (shown in Fig.~\ref{fig1}) of two single (independent) BCS bands. The same for Figures~(\ref{deltamu2}), (\ref{fig4}) and (\ref{deltamu4}).

\subsection{Determination of the Critical Chemical Potential Asymmetry $\delta \mu_{c}$ for Symmetric $V(k)$}

\begin{figure}
\centering
 \includegraphics[width=8cm]{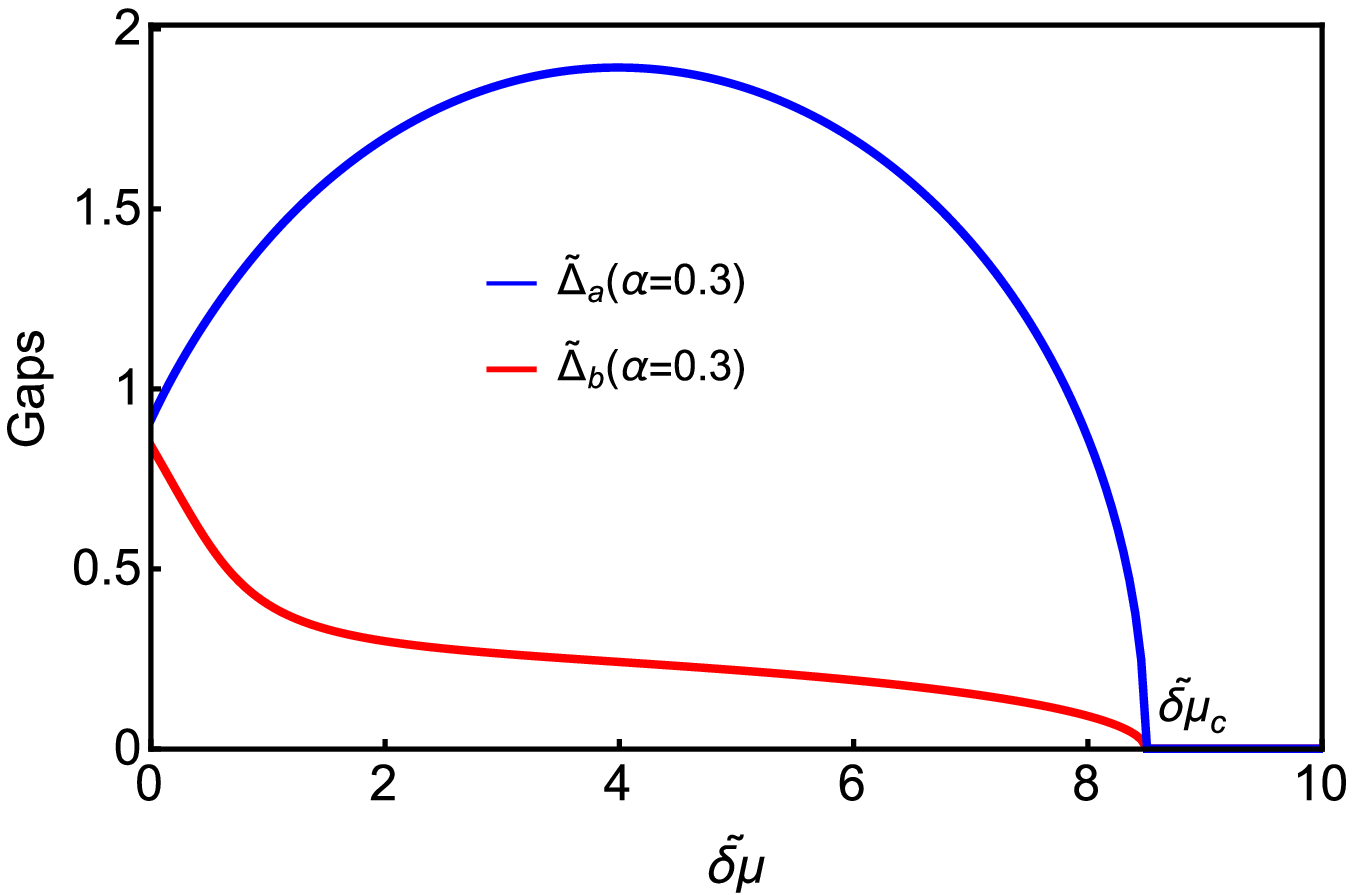}
\caption{(Color online) Pairing gaps $\tilde \Delta_a$ and $\tilde \Delta_b$ as a function of the chemical potentials asymmetry $\tilde \delta \mu$. The parameters used are $\bar \omega=10$, $\lambda_a = 0.58$, $\lambda_b = 0.6$, $\tilde \mu_a = 1.2$ and $\tilde T=0.1$.}
\label{deltamu1}
\end{figure}

In Fig.~(\ref{deltamu1}) we show $\tilde \Delta_a$ and $\tilde \Delta_b$ as a function of the chemical potential asymmetry $\tilde \delta \mu$, i.e., the asymmetry between the bottom of the $a$ and $b$ bands. It is also shown the critical chemical potential asymmetry $\tilde \delta \mu_c$ above which there are no more pairing gaps, no matter how strong the couplings $g_a$ and $g_b$ are. In the self-consistent solutions we have set $\tilde T=0.1$ and $\tilde \mu_a=1.2$.

Notice that in Fig.~(\ref{deltamu1}) there is no calculation of a critical $\tilde \delta \mu$ for $\alpha=0$, since at zero hybridization the gap equations are decoupled into two BCS gap equations for bands $a$ and $b$, which do not depend on $\tilde \delta \mu$~\cite{Twoband}.

\begin{figure}
\centering
 \includegraphics[width=8cm]{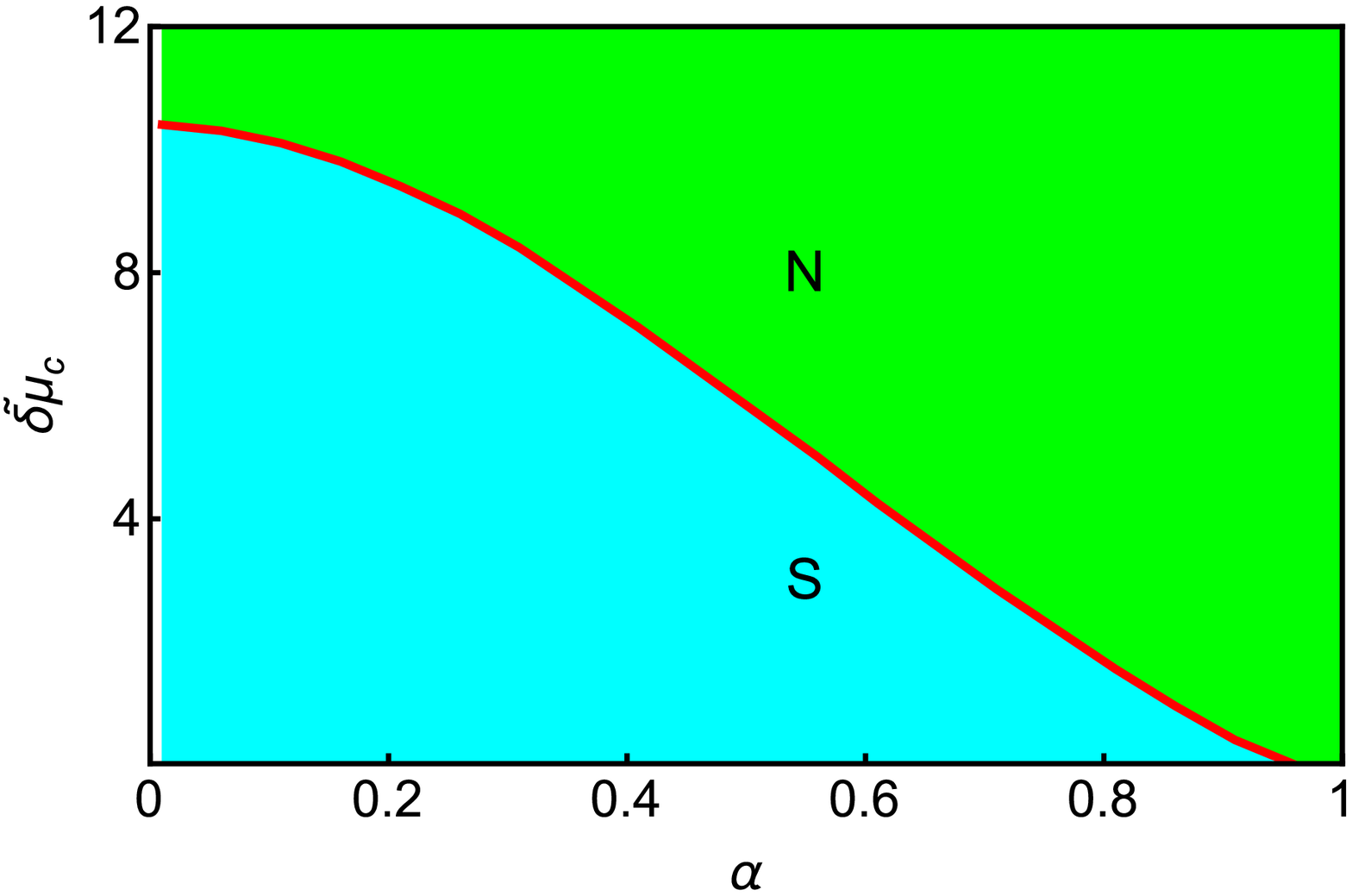}
\caption{(Color online) Phase diagram $\tilde \delta \mu_c$ versus $\alpha$ for symmetric hybridization. The parameters used are $\bar \omega=10$, $\lambda_a = 0.58$, $\lambda_b = 0.6$, $\tilde \mu_a = 1.2$ and $\tilde T=0.1$.}
\label{deltamu2}
\end{figure}

With these results we can construct the phase diagram $\tilde \delta \mu_c$ versus $\alpha$ for symmetric hybridization, which is shown in Fig.~(\ref{deltamu2}). Notice that in this situation there is no reentrant behavior, as in the previous phase diagram.

\subsection{Thermal Phase Transition as a Function of $\tilde \delta \mu$ for Symmetric $V(k)$.}

In this subsection we will find how the critical temperature $\tilde T_c$ of a two-band system, i.e., for a non-vanishing $\alpha$, behaves as a function of the chemical potential asymmetry $\tilde \delta \mu$ between the bands. With this we will be able to build the phase diagram $\tilde T_c$ versus $\tilde \delta \mu$ of the two-band model.

Since $\delta \mu$ drives phase transitions in other contexts of condensed matter~\cite{NPB,PRB}, and cold atom physics~\cite{PRL,PRA}, we expect it will play an important role here also in the normal-superconductor phase transitions.

\begin{figure}
\centering
 \includegraphics[width=8cm]{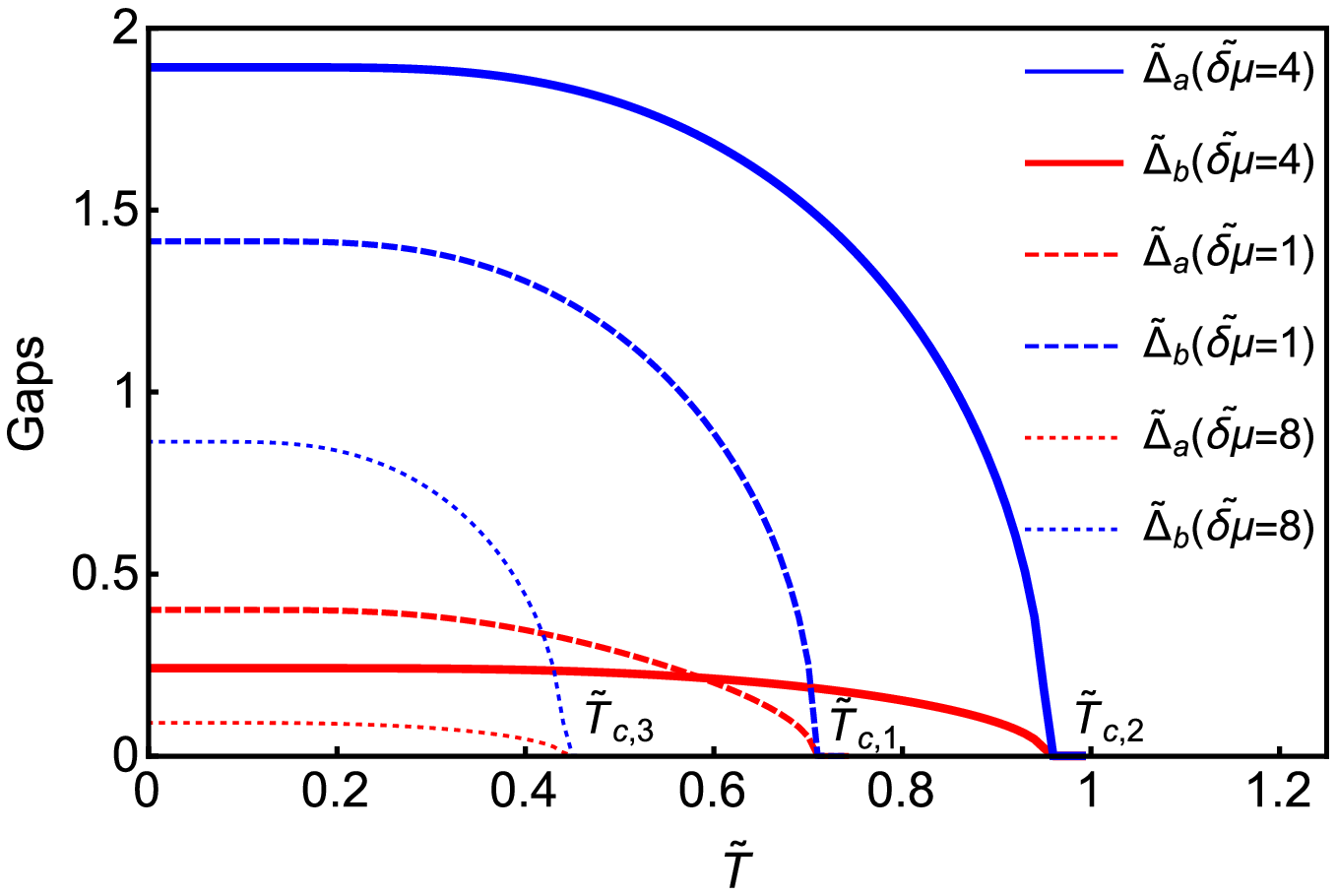}
\caption{(Color online) Gap parameters $\tilde \Delta_a$ and $\tilde \Delta_b$ versus temperature for symmetric hybridization. The parameters used are $\bar \omega=10$, $\lambda_a = 0.58$, $\lambda_b = 0.6$, $\tilde \mu_a = 1.2$, and $\alpha=0.3$.}
\label{gapsvsdeltamu1}
\end{figure}

\begin{figure}
\centering
 \includegraphics[width=8cm]{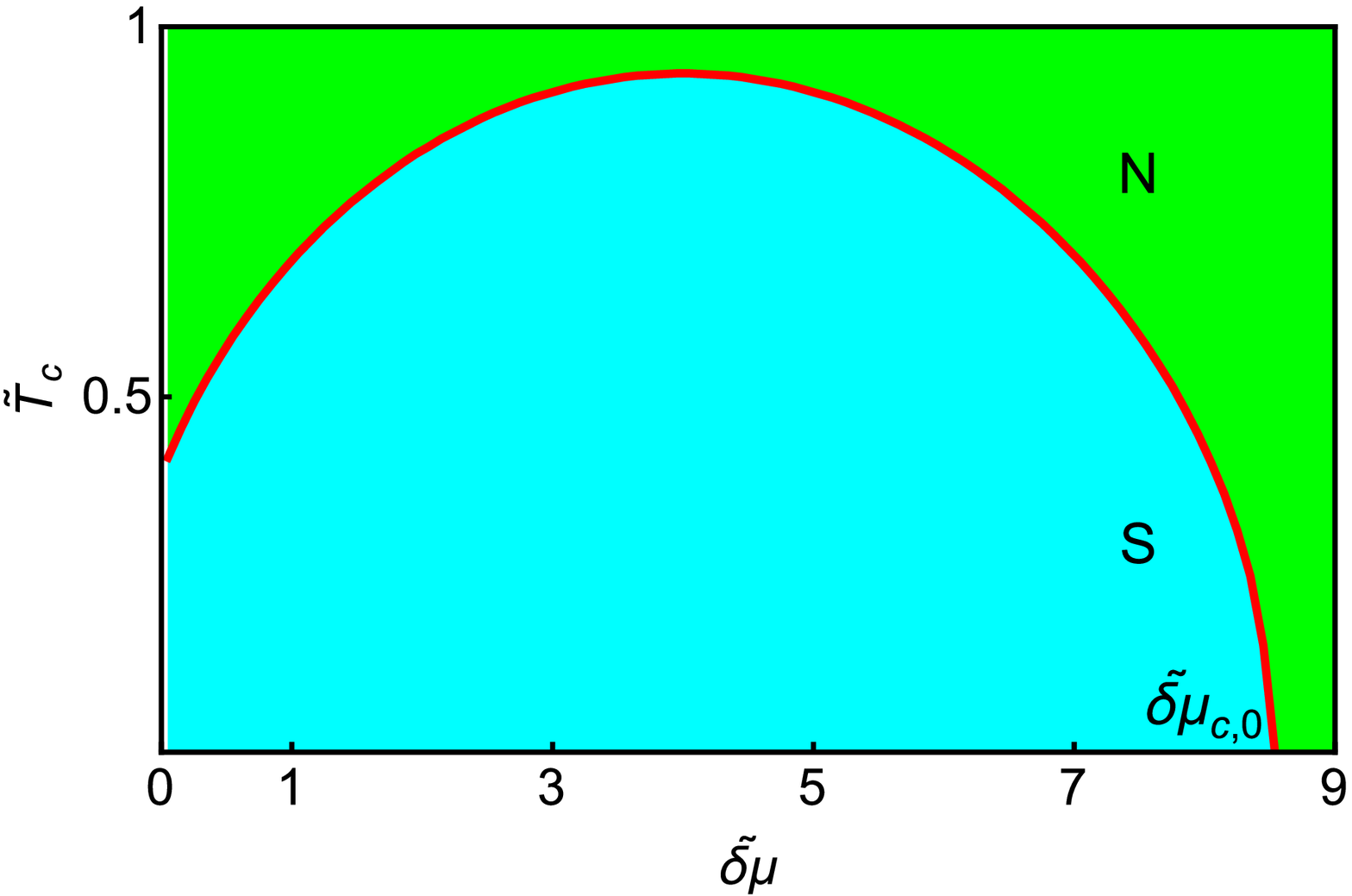}
\caption{(Color online) Phase diagram $\tilde T_c$ versus $\tilde \delta \mu$ for symmetric hybridization and $\alpha=0.3$. The parameters used are $\bar \omega=10$, $\lambda_a = 0.58$, $\lambda_b = 0.6$, $\tilde \mu_a = 1.2$.}
\label{gapsvsdeltamu2}
\end{figure}

To begin with, we plot in Fig.~(\ref{gapsvsdeltamu1}) the gap parameters with symmetric $V(k)$ as a function of the temperature $\tilde T$, for a fixed $\alpha=0.3$, and various $\tilde \delta \mu$. These curves allowed us to construct the phase diagram $\tilde T_c$ versus $\tilde \delta \mu$ of the two-band model, which is depicted in Fig.~(\ref{gapsvsdeltamu2}). The point $(\tilde \delta \mu_{c , 0},0)$ in the horizontal axis of the phase diagram is a first-order phase transition and means that coming from the right, or from a very large asymmetry where the system is deep in the normal phase, the gaps will jump from zero to $\tilde \Delta_{a,0}$ and $\tilde \Delta_{b,0}$ at $\tilde \delta \mu_{c,0}$.  In other words, for any $\tilde \delta \mu > \tilde \delta \mu_{c,0}$ the only solution is $\tilde T_c=0$, which means that $\tilde \Delta_{a} = 0$ and $\tilde \Delta_{b}=0$ in this region. This corresponds to a first-order phase transition since there is an abrupt transition from $\tilde \Delta_{a} = 0$ to $\tilde \Delta_{a} = \tilde \Delta_{a,0}$, and from $\tilde \Delta_{b}=0$ to $\tilde \Delta_{b} =\tilde \Delta_{b,0}$ at $\tilde \delta \mu_{c,0}$.


Before going to the next section, it is worth to notice that the analysis of Figs.~(\ref{fig2}) and~(\ref{gapsvsdeltamu2}) show that for optimal values of both $\alpha$ and $\tilde \delta \mu$ there is a significant enhancement of the critical temperature of the model. For optimal values of $\alpha$ and $\tilde \delta \mu$, we define those for which the first derivatives of the (increasing) functions $\tilde T_c(\alpha)$ and $\tilde T_c(\tilde \delta \mu)$ are zero. As we will se below, we arrived at the same conclusions for antisymmetric hybridization, according to Figs.~(\ref{fig4})~and (\ref{deltamu4}).

\section{\label{gap2} Obtention of the Gap Equations for Antisymmetric Hybridization $V(k)$}

We use the same procedure developed in the previous section, but now for anti-symmetric hybridization, $V_{-k} = - V_{k}$. The Hamiltonian in Eq.~(\ref{eq:Hab}) can be rewritten in the basis~$\Psi_{{\bf k}} = (a_{\textbf{k},\uparrow}, b_{\textbf{k}, \uparrow},
a_{-\textbf{k},\downarrow}^{\dag}, b_{-\textbf{k},\downarrow}^{\dag})^{T}$ as:

\begin{eqnarray}
H&=&\frac{1}{2} \sum_{{\bf k}}\Psi_{{\bf k}}^{\dag}\mathcal{H}({\bf k})\Psi_{{\bf k}}
 + 2 \sum_{\textbf{k}} \varepsilon_{+}({\bf k})\\
 \nonumber
  &-& \frac{ |\Delta_a|^{2}}{g_a} - \frac{ |\Delta_b|^{2}}{g_b}, 
 \label{6}
\end{eqnarray}
where now

\begin{eqnarray}
\mathcal{H}({\bf k})=\left(
  \begin{array}{cccc}
    \varepsilon_a ({\bf k}) & - V_{{\bf{k}}}^{*}  & -\Delta_{a}^{*} &  0  \\
 - V_{{\bf{k}}} & \varepsilon_b ({\bf k}) & 0 & -\Delta_{b}^{*} \\
   - \Delta_a & 0 & -\varepsilon_a ({\bf k}) &  -V_{{\bf{k}}} \\
    0 & -\Delta_b & -V_{{\bf{k}}}^{*} &  -\varepsilon_b ({\bf k})  \\
  \end{array}
\right). 
\end{eqnarray}

Diagonalizing the Hamiltonian (\ref{6}), we can write
\begin{eqnarray}
&&H=\sum_{\textbf{k},s=1,2}
E_{\textbf{k},s}\alpha_{\textbf{k},s}^{\dag}\alpha_{\textbf{k},s}
+ \sum_{\textbf{k},s=1,2}
(2\varepsilon_{+}({\bf k}) -E_{\textbf{k},s}) \nonumber\\
\nonumber\\
&&\;\;\;\;- \frac{|\Delta_a|^{2}}{g_a} - \frac{|\Delta_b|^{2}}{g_b}, 
\label{h2}
\end{eqnarray}
where, $\alpha_{\textbf{k}, 1,2}^{\dag}(\alpha_{\textbf{k}, 1,2})$ is the creation(annihilation) operator for the quasiparticles with excitation spectra

\begin{widetext}
\begin{eqnarray}
E_{\textbf{k}, 1,2}\!=\!\frac{1}{2} \sqrt{ 2  E_{{\bf{k}}}^{2} \!+\! 4 |V_{{\bf{k}}}|^2 \!\pm \!2 \sqrt{\! \big( |\Delta_a|^{2}\!-\!|\Delta_b|^{2} \!+\! \varepsilon_a ({\bf k})^2 \!-\! \varepsilon_b ({\bf k})^2 \big)^2\!+\! 4|V_{{\bf{k}}}|^2\Big[\big(\varepsilon_a ({\bf k}) \!+\! \varepsilon_b ({\bf k}) \big)^2 \!+\! |\Delta_{a}|^{2} \!+\! |\Delta_{b}|^{2}\Big] \!-\! 8 Re[\Delta_{a}\Delta_{b}^{*} V_{{\bf{k}}}^{2}]} }, \nonumber\\
\end{eqnarray}
where we have defined $E_{{\bf{k}}}^{2} \equiv  |\Delta_a|^{2}+|\Delta_b|^{2} + \varepsilon_a ({\bf k})^2 + \varepsilon_b ({\bf k})^2 $. The other two quasiparticles energies are $E_{\textbf{k}, 3}=-E_{\textbf{k}, 1}$ and $E_{\textbf{k}, 4}=-E_{\textbf{k}, 2}$.  We assume without loss of generality that the order parameters $\Delta_{a}$ and $\Delta_{b}$ are real. Since
the anti-symmetric hybridization $V_{{\bf{k}}}$ has to be purely imaginary to preserve time reversal symmetry~\cite{Annals1}, the term $Re[\Delta_{a}\Delta_{b}^{*} V_{{\bf{k}}}^2]$ turns out to be the same as $-\Delta_a\Delta_b|V_{{\bf{k}}}|^2$. So the previous equation can be simplified as

\begin{eqnarray}
E_{\textbf{k}, 1,2}=\frac{1}{2} \sqrt{ 2  E_{k}^{2} + 4 |V_{k}|^2 \pm 2 \sqrt{ \big( \Delta_a^{2}-\Delta_b^{2} + \varepsilon_a ({\bf k})^2 - \varepsilon_b ({\bf k})^2 \big)^2\!+\! 4|V_{{\bf{k}}}|^2\Big[\big(\varepsilon_a ({\bf k}) +\varepsilon_b ({\bf k}) \big)^2 + (\Delta_{a} + \Delta_{b})^{2}\Big]} }. \nonumber\\
\label{EnergiasAS}
\end{eqnarray}
\end{widetext}

 \subsection{\label{TP} The Grand Thermodynamic Potential for Antisymmetric Hybridization $V(k)$}

As in the previous section, the grand thermodynamic potential is written as

\begin{eqnarray}
\label{poteff1}
\Omega &=&
\frac{1}{2}\sum_{\textbf{k},s=1,2} \left[\varepsilon_{+}({\bf k})-E_{\textbf{k},s} -\frac{2}{\beta} \ln(1+e^{-\beta E_{\textbf{k},s}}) \right] \nonumber\\
&-& \frac{\Delta_a^{2}}{g_a} - \frac{\Delta_b^{2}}{g_b}.
\end{eqnarray}
The quasiparticles excitation energies $E_{\textbf{k},s}$ above are given by Eq.~(\ref{EnergiasAS}).

Minimization of $\Omega$ in Eq.~(\ref{poteff1}) with respect to the gaps $\Delta_a$ and $\Delta_b$ respectively, gives

\begin{eqnarray}
\label{gapequaG1AS}
&&\frac{4\Delta_a}{g_a} = \sum_{\textbf{k}} \! \left[\tanh \left(\frac{E_{\textbf{k},1}}{2T} \right) \frac{\partial E_{\textbf{k},1}}{\partial \Delta_a} \!+\! \tanh\left(\frac{E_{\textbf{k},2}}{2T}\right) \frac{\partial E_{\textbf{k},2}}{\partial \Delta_a}  \right], \nonumber\\
\end{eqnarray}
and
\begin{eqnarray}
\label{gapequaG2AS}
&&\frac{4\Delta_b}{g_b} \!=\! \sum_{\textbf{k}} \! \left[\tanh \left(\frac{E_{\textbf{k},1}}{2T} \right) \frac{\partial E_{\textbf{k},1}}{\partial \Delta_b} \!+ \! \tanh \left( \frac{E_{\textbf{k},2}}{2T} \right) \frac{\partial E_{\textbf{k},2}}{\partial \Delta_b}  \right]. \nonumber\\
\end{eqnarray}

As for the case of symmetric hybridizations, Eqs.~(\ref{gapequaG1AS}) and~(\ref{gapequaG2AS}) are developed in Appendix~\ref{GE2}. We define $\frac{1}{\lambda_a} = {\cal{F}}_{as}$ and $\frac{1}{\lambda_b} = {\cal{G}}_{as}$, where the subscript $as$ refers to antisymmetric. We consider a pure imaginary anti-symmetric hybridization $V_{{\bf k}}= i \gamma ({\bf k_x} k_x + {\bf k_y} k_y + {\bf k_z} k_z)$, such that $|V_k|^2 =  \gamma^2 k^2$. Using $\xi = \frac{k^2}{2m} - \bar \mu $, $|V_k|^2 \to |V_\xi|^2 = 2m \gamma^2 (\xi + \bar \mu)$. Since $[\gamma]= [k/m]$, we set $[\gamma] = [k_F/m]$, then $\gamma^2 = k_F^2/m^2$ or $m \gamma^2 = k_F^2/m = 2 E_F$. So we can define the non-dimensional hybridization parameter for anti-symmetric hybridization $\alpha = 2m\gamma^2/E_F$, such that $|V_\xi|^2 =  \alpha E_F (\xi + \bar \mu)$.

Defining again the non-dimensional variables $x=\xi/E_F$, $\tilde T = T/E_F$, $\tilde \Delta_{a,b} = \Delta_{a,b}/E_F$, $\tilde \delta \mu = \delta \mu/E_F$, $\tilde \mu = \bar \mu/E_F$, and $\tilde V_x = V_\xi/E_F$, where $E_F = k_F^2/2m$ is the Fermi energy. Defining also a non-dimensional (for the case of the antisymmetric hybridization we are considering) hybridization parameter $\alpha = 2m\gamma^2/E_F$, we have $|V_\xi|^2 =  \alpha E_F (\xi + \bar \mu) \to |V_x|^2 =  \alpha (x + \tilde \mu)$. Thus we can write

\begin{eqnarray}
\label{ASGE1}
{\cal{F}}_{as} &=& \frac{1}{4} \int_0^{\bar \omega} d x \bigg\{ \tanh \left(\frac{E_{x,1}}{2\tilde T} \right) \frac{1}{E_{x,1}} \Bigg[ 1   \\
\nonumber
&+& \left. \frac{1}{E(x)} \Big[ \tilde \Delta_a^2 - \tilde \Delta_b^2 +4 \tilde \delta \mu x + 2 |V_{{x}}|^2 \frac{\left(\tilde \Delta_a +\tilde \Delta_b \right)}{\tilde \Delta_a} \Big] \right]\bigg\} \\
\nonumber
&+& \frac{1}{4} \int_0^{\bar \omega} d x \bigg\{  \tanh \left(\frac{E_{x,2}}{2\tilde T} \right) \frac{1}{E_{x,2}} \Bigg[ 1   \\
\nonumber
&-& \left. \frac{1}{E(x)} \Big[ \tilde \Delta_a^2 - \tilde \Delta_b^2 + 4 \tilde \delta \mu x  + 2 |V_{{x}}|^2 \frac{\left(\tilde \Delta_a + \tilde \Delta_b \right)}{\tilde \Delta_a} \Big] \right] \bigg\},
\end{eqnarray}
and

\begin{eqnarray}
\label{ASGE2}
{\cal{G}}_{as} & =& \frac{1}{4} \int_0^{\bar \omega} d \xi \bigg\{ \tanh \left(\frac{E_{x,1}}{2 \tilde T} \right) \frac{1}{E_{x,1}} \Bigg[ 1  \\
\nonumber
&-& \left. \frac{1}{E(x)} \Big[  \tilde \Delta_a^2 - \tilde \Delta_b^2 + 4 \delta \mu x + 2 |V_{{x}}|^2 \frac{\left(\tilde \Delta_a + \tilde \Delta_b \right)}{\tilde \Delta_b} \Big] \right]\bigg\} \\
\nonumber
&+& \frac{1}{4} \int_0^{\bar \omega} d \xi \bigg\{ \tanh \left(\frac{E_{x,2}}{2 \tilde T} \right) \frac{1}{E_{x,2}} \Bigg[ 1   \\
\nonumber
&+& \left. \frac{1}{E(x)} \Big[  \tilde \Delta_a^2 - \tilde \Delta_b^2 + 4 \delta \mu x + 2 |V_{{x}}|^2 \frac{\left( \tilde \Delta_a + \tilde \Delta_b \right)}{\tilde \Delta_b} \Big] \right] \bigg\},
\end{eqnarray}
where $\bar \omega=\omega/E_F$, $E_{{x}, 1,2}=\frac{1}{2} \sqrt{ 2  E_{x}^{2} + 4 \tilde |V_{x}|^2 \pm 2  E(x)}$, $E_{x}^{2} =  \tilde \Delta_a^{2}+\tilde \Delta_b^{2} + 2 ({x}^2+{\tilde \delta \mu}^2)$ and $E(x)=\sqrt{ \big(\tilde \Delta_a^{2}- \tilde \Delta_b^{2} + 4\tilde \delta \mu x \big)^2+ 4 \tilde |V_{x}|^2\Big[( 2 x )^2 + (\tilde \Delta_{a} + \tilde \Delta_{b})^{2} \Big] }$.

\subsection{Determination of the Critical Temperatures $T_{c,a}$ and $T_{c,b}$ for Antisymmetric $V(k)$}

\begin{figure}
\centering
 \includegraphics[width=8cm]{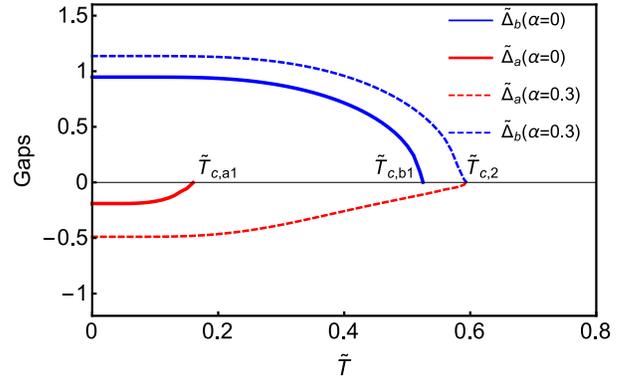}
\caption{(Color online) Gap parameters with antisymmetric hybridization as a function of the temperature for two values of $\alpha=2m\gamma^2/E_F$ for $\bar \omega=10$, $\lambda_a = 0.58$, $\lambda_b = 0.6$, $\tilde \mu_a = 1.2$, and $\tilde \mu_b = 1.6.$}
\label{fig3}
\end{figure}

As for the case of symmetric hybridization, given $g_a$ and $g_b$, Eqs.~(\ref{ASGE1}) and (\ref{ASGE2}) have to be solved self-consistently to find the gaps $\tilde \Delta_a$ and $\tilde \Delta_b$. In Fig.~(\ref{fig3}) the normalized gaps $\tilde \Delta_a$ and $\tilde \Delta_b$ are shown as a function of the temperature for two values of the hybridization parameter $\alpha=2m\gamma^2/E_F$ for the antisymmetric hybridization we are considering. The curves are the (self-consistent) solutions of Eqs.~(\ref{AntiSym1}) and (\ref{AntiSym2}). 

\begin{figure}
\centering
 \includegraphics[width=8cm]{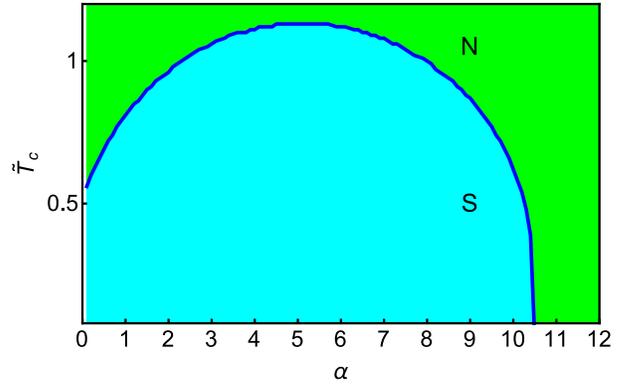}
\caption{(Color online) Critical temperature $\tilde T_c$ versus $\alpha$ for antisymmetric hybridization. The parameters used are for $\bar \omega=10$, $\lambda_a = 0.58$, $\lambda_b = 0.6$, $\tilde \mu_a = 1.2$, and $\tilde \mu_b = 1.6.$}
\label{fig4}
\end{figure}

As happened for the case of symmetric hybridization, Fig.~(\ref{fig3}) shows that for any $\alpha>0$ there is only one critical temperature where both gaps vanish. This allows us to construct a finite temperature phase diagram of this two-band model with antisymmetric hybridization, which is depicted In Fig.~(\ref{fig4}). Again, this phase diagram evidences reentrant behavior.

\subsection{Determination of the Critical Chemical Potential Asymmetry $\delta \mu_{c}$ for Antisymmetric $V(k)$}

\begin{figure}
\centering
 \includegraphics[width=8cm]{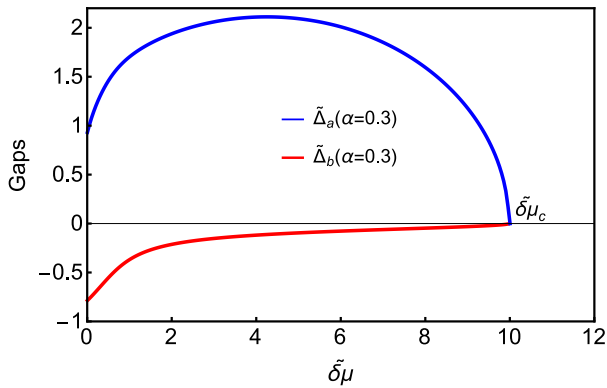}
\caption{(Color online) Pairing gaps $\Delta_a$ and $\Delta_b$ as a function of the chemical potentials asymmetry $\tilde \delta \mu$. The parameters used are $X=10$, $\lambda_a = 0.58$, $\lambda_b = 0.6$, $\tilde \mu_a = 1.2$ and $\tilde T=0.1$.}
\label{deltamu3}
\end{figure}

In Fig.~(\ref{deltamu3}) we show the critical chemical potential asymmetry above which there are no more pairing gaps $\tilde \Delta_a$ and $\tilde \Delta_b$, no matter how strong the couplings $g_a$ and $g_b$ are. In the self-consistent solutions we have set $\tilde T=0.1$ and $\tilde \mu_a=1.2$.

\begin{figure}
\centering
 \includegraphics[width=8cm]{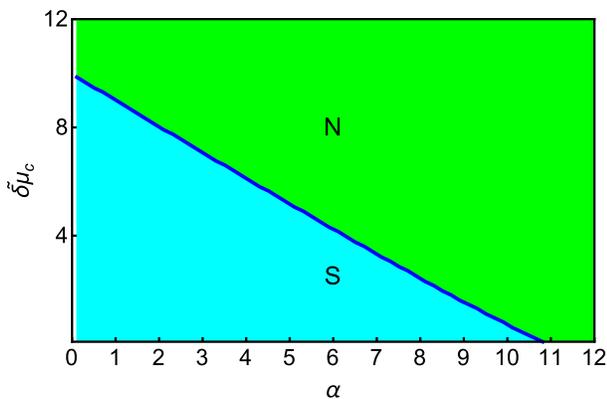}
\caption{(Color online) Phase diagram $\tilde \delta \mu_c$ versus $\alpha$ for antisymmetric hybridization. The parameters used are $X=10$, $\lambda_a = 0.58$, $\lambda_b = 0.6$, $\tilde \mu_a = 1.2$ and $\tilde T=0.1$.}
\label{deltamu4}
\end{figure}

The above results were used to construct the phase diagram $\tilde \delta \mu_c$ versus $\alpha$ for antisymmetric hybridization, which is plotted in Fig.~(\ref{deltamu4}).

\subsection{Thermal Phase Transition as a Function of $\tilde \delta \mu$ for Antisymmetric $V(k)$.}

In this subsection we will find how the critical temperature $\tilde T_c$ of a two-band system, i.e., for a non-vanishing $\alpha$, behaves as a function of the chemical potential asymmetry $\tilde \delta \mu$ between the bands. With this we will be able to build the phase diagram $\tilde T_c$ versus $\tilde \delta \mu$ of the two-band model.

\begin{figure}
\centering
 \includegraphics[width=8cm]{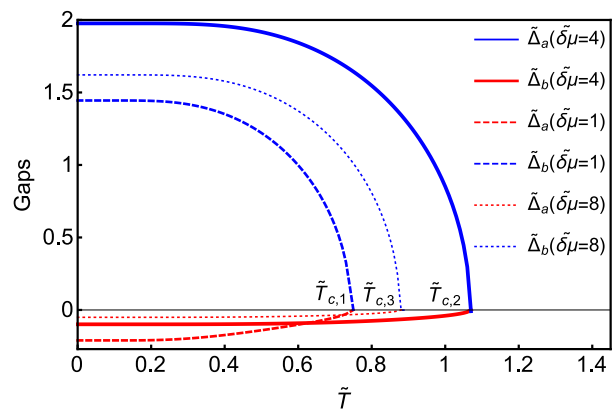}
\caption{(Color online) Gap parameters $\tilde \Delta_a$ and $\tilde \Delta_b$ versus temperature for antisymmetric hybridization. The parameters used are $X=10$, $\lambda_a = 0.58$, $\lambda_b = 0.6$, $\tilde \mu_a = 1.2$, and $\alpha=0.3$.}
\label{gapsvsdeltamu3}
\end{figure}

\begin{figure}
\centering
 \includegraphics[width=8cm]{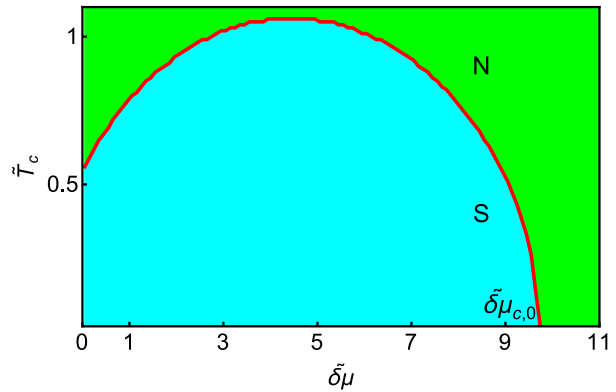}
\caption{(Color online) Phase diagram $\tilde T_c$ versus $\tilde \delta \mu$ for antisymmetric hybridization and $\alpha=0.3$. The parameters used are $X=10$, $\lambda_a = 0.58$, $\lambda_b = 0.6$, $\tilde \mu_a = 1.2$.}
\label{gapsvsdeltamu4}
\end{figure}

We plot in Fig.~(\ref{gapsvsdeltamu3}) the gap parameters with antisymmetric hybridization as a function of the temperature $\tilde T$, for a fixed $\alpha=0.3$, and various $\tilde \delta \mu$. This curves allowed us to construct the phase diagram $\tilde T_c$ versus $\tilde \delta \mu$ of the two-band model, which is depicted in Fig.~(\ref{gapsvsdeltamu4}). The point $(\tilde \delta \mu_{c,0},0)$ in the horizontal axis of the phase diagram is a first-order (quantum) phase transition. As we mentioned before, this means that coming from a very large asymmetry (from the right), where the system is deep in the normal phase, the gaps will jump from zero to $\tilde \Delta_{a,0}$ and $\tilde \Delta_{b,0}$ at $\tilde \delta \mu_{c,0}$. 

It should be noted from Figs.~(\ref{fig3}), (\ref{deltamu3}) and (\ref{gapsvsdeltamu3}) that the $a$ and $b$ gaps have different signs for any $\tilde T>0$ or $\tilde \delta \mu>0$. This behavior resembles that of iron based (IB) superconductors (SC), which are characterized by the $s_{\pm}$ gap symmetry, in which a gap changes its sign between pockets of the Fermi surface~\cite{Hanaguri}. As pointed out in Ref.~\cite{Hanaguri}, the sign reversal in the SC-gap function implies that the electron pairing is mediated by repulsive interaction in $k$ space, which is supposedly spin fluctuations, associated with the nesting between the Fermi pockets. Here, the sign reversal in the SC-gaps could be qualitatively understood as a consequence of the change of the two Fermi surfaces due to the antisymmetric hybridization between them.

\section{\label{Co} Summary and Conclusions}

We have investigated the temperature effects on the superconducting properties of a two-band system with symmetric and anti-symmetric hybridization $V(k)$. We considered that these bands are formed by electronic orbitals with angular momenta, such that their hybridization can be symmetric or anti-symmetric under inversion symmetry. We have taken into account only intra-band attractive interactions in the two bands, responsible for intra-band $s$-wave pairing gaps $\Delta_a$ and $\Delta_b$, and investigated how the critical temperature $T_c$ of the system depend on the strength of the hybridization $\alpha$ and on the chemical potential asymmetry $\delta \mu$ between the bottom of the two-bands. We have also constructed the phase diagrams of the two-band model $T_c$ versus $\alpha$ and $T_c$ versus $\delta \mu$ for both symmetric and antisymmetric hybridizations.

We have seen in the phase diagrams $T_c$ versus $\alpha$ and $T_c$ versus $\tilde \delta \mu$, for symmetric and antisymmetric hybridizations, that the two-band model presents a type of ``reentrant'' behavior (RB). In the RB, the system is initially in the normal phase $N$ and as the parameters $\alpha$ or $\tilde \delta \mu$ (which drive the phase transitions) increase, the system enters the superconducting phase $S$ and then enters again in the normal phase $N$.

We have also shown that for any $\alpha > 0$ there is only one $T_c$ for the system, where both gaps vanish simultaneously. This fact makes this simple generic two-band model at least a laboratory for studying real materials, since it reproduces, for instance, the same (albeit in a qualitative way) behavior found experimentally for ${\rm MgB_2}$.

We have also shown that for optimal values of both $\alpha$ and $\tilde \delta \mu$ there is a significant enhancement of the critical temperature of the model, for symmetric and antisymmetric hybridization. Since hybridization can be done by doping or pressure, the enhancement of the critical temperature and, consequently, the reentrant behavior could be seen experimentally, provided the material is satisfactorily described by the hybridized two-band model investigated here.

\section{\label{Ac} Acknowledgments}

We wish to thank the Brazilian agencies CAPES and CNPq for financial support.

\appendix
\section{Derivation of the Gap Equations For Symmetric Hybridization}
\label{GE1}

For symmetric hybridization, Eqs.~(\ref{gapequaG1}) and (\ref{gapequaG2}) give

\begin{eqnarray}
\label{gapequaT1}
&&\frac{\Delta_a}{g_a} = \frac{1}{8} \int \frac{d k~k^2}{2 \pi^2} \tanh \left(\frac{E_{\textbf{k},1}}{2T} \right) \frac{1}{E_{\textbf{k},1}} \Bigg[ \Delta_a  + \\
\nonumber
&& \left. \frac{1}{E(k)} \Big[ \Delta_a \left(\Delta_a^2 - \Delta_b^2 + \varepsilon_a ({\bf k})^2 - \varepsilon_b ({\bf k})^2 \right) + 2 V_{{\bf{k}}}^2 \left( \Delta_a - \Delta_b \right) \Big] \right]\\
\nonumber
&+& \tanh \left(\frac{E_{\textbf{k},2}}{2T} \right) \frac{1}{E_{\textbf{k},2}} \Bigg[ \Delta_a -  \\
\nonumber
&& \left. \frac{1}{E(k)} \Big[ \Delta_a \left(\Delta_a^2 - \Delta_b^2 + \varepsilon_a ({\bf k})^2 - \varepsilon_b ({\bf k})^2 \right) + 2 V_{{\bf{k}}}^2 \left( \Delta_a - \Delta_b \right) \Big] \right],
\end{eqnarray}
and

\begin{eqnarray}
\label{gapequaT2}
&&\frac{\Delta_b}{g_b} = \frac{1}{8} \int \frac{d k~k^2}{2 \pi^2} \tanh \left(\frac{E_{\textbf{k},1}}{2T} \right) \frac{1}{E_{\textbf{k},1}} \Bigg[ \Delta_b - \\
\nonumber
&& \left. \frac{1}{E(k)} \Big[ \Delta_b \left(\Delta_a^2 - \Delta_b^2 + \varepsilon_a ({\bf k})^2 - \varepsilon_b ({\bf k})^2 \right) + 2 V_{{\bf{k}}}^2 \left( \Delta_a - \Delta_b \right) \Big] \right]\\
\nonumber
&+& \tanh \left(\frac{E_{\textbf{k},2}}{2T} \right) \frac{1}{E_{\textbf{k},2}} \Bigg[ \Delta_b  + \\
\nonumber
&& \left. \frac{1}{E(k)} \Big[ \Delta_b \left(\Delta_a^2 - \Delta_b^2 + \varepsilon_a ({\bf k})^2 - \varepsilon_b ({\bf k})^2 \right) + 2 V_{{\bf{k}}}^2 \left( \Delta_a - \Delta_b \right) \Big] \right],
\end{eqnarray}
where $E(k) \equiv \Big\{\left(\Delta_a^{2}-\Delta_b^{2} + \varepsilon_a ({\bf k})^2 - \varepsilon_b ({\bf k})^2 \right)^2 + 4V_{{\bf{k}}}^2\left[\left(\Delta_a-\Delta_b \right)^{2} + \left(\varepsilon_a ({\bf k}) + \varepsilon_b ({\bf k})\right)^2\right] \Big\}^{1/2} = E_{\textbf{k},1}^2 - E_{\textbf{k},2}^2$, with $E_{\textbf{k},1,2}$ given by~Eq.~(\ref{Energias}).

In order to integrate the equations above we verified that after some simple algebra the momentum dependent terms in the above equations can be written as

\begin{eqnarray}
\label{simple}
\varepsilon_a ({\bf k}) + \varepsilon_b ({\bf k}) &=& 2 {\xi_k}\\
\nonumber
\varepsilon_a ({\bf k})^2 - \varepsilon_b ({\bf k})^2 &=& 4 \delta \mu \xi_k,\\
\nonumber
\varepsilon_a ({\bf k})^2 + \varepsilon_b ({\bf k})^2 &=& 2 [{\xi_k}^2+{\delta \mu}^2],
\nonumber
\end{eqnarray}
where we have defined $\xi_k = \frac{k^2}{2m} - \bar \mu $, $\bar \mu = (\mu_a + \mu_b)/2=\mu + (E_a + E_b)/2$ and $\delta \mu = (\mu_b - \mu_a)/2 = (E_b - E_a)/2$. This will allow us to make the appropriate (usual) change of variables $\xi = \frac{k^2}{2m} - \bar \mu $ to proceed with the integration of the gap equations,

\begin{eqnarray}
\label{gapequaT3}
&&\frac{\Delta_a}{\lambda_a} = \frac{1}{4} \int_0^{\omega} d \xi \bigg\{ \tanh \left(\frac{E_{\xi,1}}{2T} \right) \frac{1}{E_{\xi,1}} \Bigg[ \Delta_a   \\
\nonumber
&+& \left. \frac{1}{E(\xi)} \Big[ \Delta_a \left(\Delta_a^2 - \Delta_b^2 +4 \delta \mu \xi \right) + 2 V_{{\xi}}^2 \left( \Delta_a - \Delta_b \right) \Big] \right]\\
\nonumber
&+& \tanh \left(\frac{E_{\xi,2}}{2T} \right) \frac{1}{E_{\xi,2}} \Bigg[ \Delta_a   \\
\nonumber
&-& \left. \frac{1}{E(\xi)} \Big[ \Delta_a \left(\Delta_a^2 - \Delta_b^2 + 4 \delta \mu \xi \right) + 2 V_{{\xi}}^2 \left( \Delta_a - \Delta_b \right) \Big] \right] \bigg\},
\end{eqnarray}
and

\begin{eqnarray}
\label{gapequaT4}
&&\frac{\Delta_b}{\lambda_b} = \frac{1}{4} \int_0^{\omega} d \xi \bigg\{ \tanh \left(\frac{E_{\xi,1}}{2T} \right) \frac{1}{E_{\xi,1}} \Bigg[ \Delta_b  \\
\nonumber
&-& \left. \frac{1}{E(\xi)} \Big[ \Delta_b \left(\Delta_a^2 - \Delta_b^2 + 4 \delta \mu \xi \right) + 2 V_{{\xi}}^2 \left( \Delta_a - \Delta_b \right) \Big] \right]\\
\nonumber
&+& \tanh \left(\frac{E_{\xi,2}}{2T} \right) \frac{1}{E_{\xi,2}} \Bigg[ \Delta_b   \\
\nonumber
&+& \left. \frac{1}{E(\xi)} \Big[ \Delta_b \left(\Delta_a^2 - \Delta_b^2 + 4 \delta \mu \xi \right) + 2 V_{{\xi}}^2 \left( \Delta_a - \Delta_b \right) \Big] \right] \bigg\},
\end{eqnarray}
where $\lambda_a \equiv g_a \rho(0)$, $\lambda_b \equiv g_b \rho(0)$, with $\rho(0) = \frac{m}{2 \pi^2} k_F$ and $\omega$ is an energy cutoff. Here $k_F=\sqrt{2m \bar \mu}$ is the Fermi momentum.  In the above equations we have taken a symmetric hybridization $V_k = \gamma k^2$, where $\gamma$ is the strength of the hybridization, and $k^2 = k_x^2+ k_y^2+ k_z^2$, such that $V(-k)=V(k)$. Thus, $V_{\xi} = 2m \gamma (\bar \mu + \xi)$. The quasiparticle energies now read

\begin{widetext}
\begin{eqnarray}
\label{EnergiasNew}
E_{{\xi}, 1,2}=\frac{1}{2} \sqrt{ 2  E_{\xi}^{2} + 4 V_{\xi}^2 \pm 2 \sqrt{ \big( \Delta_a^{2}-\Delta_b^{2} + 4 \delta \mu \xi \big)^2+ 4V_{\xi}^2\Big[\big( 2 \xi \big)^2 + \big(\Delta_{a} - \Delta_{b}\big)^{2} \Big] } }. \nonumber\\
\end{eqnarray}
\end{widetext}
Besides, $E_{\xi}^{2} =  \Delta_a^{2}+\Delta_b^{2} + 2 [{\xi}^2+{\delta \mu}^2] $, and $E(\xi) = \Big\{\left(\Delta_a^{2}-\Delta_b^{2} + 4 \delta \mu \xi \right)^2 + 4V_{\xi}^2\left[\left(\Delta_a-\Delta_b \right)^{2} + \left( 2 \xi \right)^2\right] \Big\}^{1/2}$. 

We can write Eqs.~(\ref{gapequaT3}) and (\ref{gapequaT4}) as

\begin{equation}
\label{PreGL1}
\frac{1}{\lambda_a} = {\cal{F}}(\Delta_a,\Delta_b,\gamma) \equiv {\cal{F}},
\end{equation}

\begin{equation}
\label{PreGL2}
\frac{1}{\lambda_b} = {\cal{G}}(\Delta_a,\Delta_b,\gamma)  \equiv {\cal{G}},
\end{equation}
where

\begin{eqnarray}
\label{PreGL3}
{\cal{F}} &=& \frac{1}{4} \int_0^{\omega} d \xi \bigg\{ \tanh \left(\frac{E_{\xi,1}}{2T} \right) \frac{1}{E_{\xi,1}} \Bigg[ 1   \\
\nonumber
&+& \left. \frac{1}{E(\xi)} \Big[ \left(\Delta_a^2 - \Delta_b^2 +4 \delta \mu \xi \right) + 2 V_{{\xi}}^2 \frac{\left( \Delta_a - \Delta_b \right)}{\Delta_a} \Big] \right]\bigg\} \\
\nonumber
&+& \frac{1}{4} \int_0^{\omega} d \xi \bigg\{  \tanh \left(\frac{E_{\xi,2}}{2T} \right) \frac{1}{E_{\xi,2}} \Bigg[ 1   \\
\nonumber
&-& \left. \frac{1}{E(\xi)} \Big[ \left(\Delta_a^2 - \Delta_b^2 + 4 \delta \mu \xi \right) + 2 V_{{\xi}}^2 \frac{\left( \Delta_a - \Delta_b \right)}{\Delta_a} \Big] \right] \bigg\},
\end{eqnarray}
and

\begin{eqnarray}
\label{PreGL4}
{\cal{G}}& =& \frac{1}{4} \int_0^{\omega} d \xi \bigg\{ \tanh \left(\frac{E_{\xi,1}}{2T} \right) \frac{1}{E_{\xi,1}} \Bigg[ 1  \\
\nonumber
&-& \left. \frac{1}{E(\xi)} \Big[  \left(\Delta_a^2 - \Delta_b^2 + 4 \delta \mu \xi \right) + 2 V_{{\xi}}^2 \frac{\left( \Delta_a - \Delta_b \right)}{\Delta_b} \Big] \right]\bigg\} \\
\nonumber
&+& \frac{1}{4} \int_0^{\omega} d \xi \bigg\{ \tanh \left(\frac{E_{\xi,2}}{2T} \right) \frac{1}{E_{\xi,2}} \Bigg[ 1   \\
\nonumber
&+& \left. \frac{1}{E(\xi)} \Big[  \left(\Delta_a^2 - \Delta_b^2 + 4 \delta \mu \xi \right) + 2 V_{{\xi}}^2 \frac{\left( \Delta_a - \Delta_b \right)}{\Delta_b} \Big] \right] \bigg\}.
\end{eqnarray}

\section{Derivation of the Gap Equations For Antisymmetric Hybridization}
\label{GE2}

Minimizing the thermodynamic potential an making the same change of variables, as we did in the case of symmetric hybridization, we obtain $E_{{\xi}, 1,2}=\frac{1}{2} \sqrt{ 2  E_{\xi}^{2} + 4 |V_{\xi}|^2 \pm 2  E(\xi)}$, $E_{\xi}^{2} =  \Delta_a^{2}+\Delta_b^{2} + 2 ({\xi}^2+{\delta \mu}^2)$ and $E(\xi)=\sqrt{ \big( \Delta_a^{2}-\Delta_b^{2} + 4 \delta \mu \xi \big)^2+ 4|V_{\xi}|^2\Big[( 2 \xi )^2 + (\Delta_{a} + \Delta_{b})^{2} \Big] }$.

As we mentioned before, we take a pure imaginary anti-symmetric hybridization $V_{{\bf k}}= i \gamma ({\bf k_x} k_x + {\bf k_y} k_y + {\bf k_z} k_z)$, such that $|V_k|^2 =  \gamma^2 k^2 \to  2m \gamma^2 (\xi + \bar \mu)$. Since $[\gamma]= [k/m]$, we set $[\gamma] = [k_F/m]$, then $\gamma^2 = k_F^2/m^2$ or $m \gamma^2 = k_F^2/m = 2 E_F$. So we can define the non-dimensional hybridization parameter for anti-symmetric hybridization $\alpha = 2m\gamma^2/E_F$, such that $|V_\xi|^2 =  \alpha E_F (\xi + \bar \mu)$. Then, as we did in the case of symmetric hybridization, the gap equations are written as

\begin{eqnarray}
\label{AntiSym1}
&&\frac{\Delta_a}{\lambda_a} = \frac{1}{4} \int_0^{\omega} d \xi \bigg\{ \tanh \left(\frac{E_{\xi,1}}{2T} \right) \frac{1}{E_{\xi,1}} \Bigg[ \Delta_a   \\
\nonumber
&+& \left. \frac{1}{E(\xi)} \Big[ \Delta_a \left(\Delta_a^2 - \Delta_b^2 +4 \delta \mu \xi \right) + 2 |V_{{\xi}}|^2 \left( \Delta_a + \Delta_b \right) \Big] \right]\\
\nonumber
&+& \tanh \left(\frac{E_{\xi,2}}{2T} \right) \frac{1}{E_{\xi,2}} \Bigg[ \Delta_a   \\
\nonumber
&-& \left. \frac{1}{E(\xi)} \Big[ \Delta_a \left(\Delta_a^2 - \Delta_b^2 + 4 \delta \mu \xi \right) + 2 |V_{{\xi}}|^2 \left( \Delta_a + \Delta_b \right) \Big] \right] \bigg\}.
\end{eqnarray}
and

\begin{eqnarray}
\label{AntiSym2}
&&\frac{\Delta_b}{\lambda_b} = \frac{1}{4} \int_0^{\omega} d \xi \bigg\{ \tanh \left(\frac{E_{\xi,1}}{2T} \right) \frac{1}{E_{\xi,1}} \Bigg[ \Delta_b  \\
\nonumber
&-& \left. \frac{1}{E(\xi)} \Big[ \Delta_b \left(\Delta_a^2 - \Delta_b^2 + 4 \delta \mu \xi \right) + 2 |V_{{\xi}}|^2 \left( \Delta_a + \Delta_b \right) \Big] \right]\\
\nonumber
&+& \tanh \left(\frac{E_{\xi,2}}{2T} \right) \frac{1}{E_{\xi,2}} \Bigg[ \Delta_b   \\
\nonumber
&+& \left. \frac{1}{E(\xi)} \Big[ \Delta_b \left(\Delta_a^2 - \Delta_b^2 + 4 \delta \mu \xi \right) + 2 |V_{{\xi}}|^2 \left( \Delta_a + \Delta_b \right) \Big] \right] \bigg\}.
\end{eqnarray}


\begin{thebibliography}{41}
\bibitem{Comment} Strictly speaking a phase diagram is a graphical representation used to show physical quantities as temperature, pressure, volume, etc. at which thermodynamically distinct phases occur and coexist at equilibrium. In Fig.~(\ref{fig2}) (as well as in Figs.~(\ref{deltamu2}, \ref{gapsvsdeltamu2}, \ref{fig4}, \ref{deltamu4}, \ref{gapsvsdeltamu4})), what we have are lines showing the correspondent critical properties.
\end{thebibliography}
\end{document}